# End-fire injection of guided light into optical microcavity

Shuai Liu[1], Zhiyuan Gu[1], Nan Zhang[1], Kaiyang Wang[1], Shumin Xiao[2], Quan Lyu[1], and Qinghai Song[1,3,*]

**Abstract** Coupling light into microdisk plays a key role in a number of applications such as resonant filters and optical sensors. While several approaches have successfully coupled light into microdisk efficiently, most of them suffer from the ultrahigh sensitivity to the environmental vibration. Here we demonstrate a robust mechanism, which is termed as "end-fire injection". By connecting an input waveguide to a circular microdisk directly, the mechanism shows that light can be efficiently coupled into optical microcavity. The coupling efficiency can be as high as 75% when the input signals are on resonances. Our numerical results reveal that the high coupling efficiency is attributed to the constructive interference between the whispering gallery modes and the input signals. We have also shown that the end-fire injection can be further extended to the long-lived resonances with low refractive index such as $n = 1.45$. We believe our results will shed light on the applications of optical microcavities.

## 1. Introduction

Optical microcavities have attracted considerable research attentions in past decades. Whispering gallery modes (WGMs) based microcavities [1] are prominent examples. The light is confined by total internal reflection (TIR) along the cavity boundary for a long time and thus gives extremely high quality factors (Q ~ $\omega\tau$, $\omega$ is the angular frequency and $\tau$ is the lifetime). The recorded high Q factors are on the order of $10^{10}$ for crystalline resonator [2], $10^9$ for silica microsphere [3], $10^8$ for silica microtoroid [4], and $10^6$ for semiconductor microdisk [5]. Such high Q factors promise a number of applications of WGMs-based microcavities, especially in the area of distributed sensing [6-9].

For most of the applications in sensors and optical pumping

\* Qinghai Song
   Qinghai.song@hitsz.edu.cn
1 Department of Electronic and Information Engineering, Shenzhen Graduate School, Harbin Institute of Technology, Shenzhen, 518055, China
2 Department of Material Science and Engineering, Shenzhen Graduate School, Harbin Institute of Technology, Shenzhen, 518055, China
3 The National Key Laboratory on Tunable Laser Technology, Harbin Institute of Technology, Harbin, 158010, China

microdisks, efficiently coupling light into the resonators is of the central importance. Several methods have been developed in past decades and high coupling efficiencies have been reported. The prism or tapered-fiber coupling is commonly used and can give a coupling efficiency as high as 99.9% when the critical coupling condition is satisfied [10, 11]. Based on this coupling mechanism, Raman laser with ultralow threshold [1] and sensors [7, 8] have been successfully demonstrated in microspheres and microtoroids. As the time reversal process, the WGMs of microcavity can also be excited by the focused light in free space, which is known as free space coupling [12, 13]. The free space coupling does not require precise position control and the coupling efficiency can be higher than 30% [14-17]. This method has been successfully applied in biosensors and laser pumping [9, 17]. However, the above two schemes have obvious limitation in applications. For the evanescent coupling, the refractive index of microcavity must be close to the input waveguide or fiber to ensure the phase matching. The tapered fiber based coupling requires high stabilization and thus the devices are very sensitive to environmental vibrations. For the case of on-chip waveguide coupling systems, the separation distance is usually around a few hundred nanometers, which is hard to be fabricated with standard photolithography. While the vertical coupling has been demonstrated, it also requires very precise control of the overlay between waveguide and microcavity. Compared with evanescent coupling, the coupling efficiency of free space coupling is relatively low and it must be operated under the optical microscope, especially for wavelength scaled microcavity [18]. Therefore, a new coupling mechanism which can address the above limitations is highly desirable. Here we demonstrate a new scheme to couple light from waveguide to microdisk efficiently. By connecting a waveguide to microdisk directly, we show that the coupling efficiency can be as high as 75% when the input signals are on resonance. As there is no separation distance between waveguide and microdisk, the coupling system is naturally insensitive to environment vibration and also easy to be fabricated with standard photolithography. Our numerical results reveal that the surprising high efficiency is induced by the constructive interference between the input light and resonant modes.

## 2. Results and Discussion

In this letter, we treat the microdisk based system as two-dimensional objects in transverse plane by applying effective refractive index $n$. The vertical radiation loss is ignored. The simplest structure for direct coupling between microcavity and waveguide is shown in the inset of Fig. 1(a). It is a circular cavity with radius $R$, which is connected with a waveguide with width $w$. We assume the refractive index of microdisk and waveguide as $n = 3.3$ (for the material of GaAs or Silicon). The outside is air with $n = 1$. In our simulation modal, the Perfect Matched Layer (PML) and scattering boundary condition are applied to absorb the outgoing waves. The coupling between microdisk and waveguide is widely studied, usually in the way from microdisk to waveguide. Due to the breaking of TIR at the joint position, the leakage along the waveguide is orders of magnitude higher than the evanescent escape and thus dominates the far field patterns (see Fig. 1(a) for an example). Such directional outputs have been theoretically and experimentally studied in past few years, both in circular and deformed microcavities [19-23]. However, the reversed process, which is the coupling from waveguide to microdisk, has not been well explored so far. Intuitively, the efficient coupling from waveguide to microdisk seems to be impossible. While the input signals are divergent by the diffraction after they leave the waveguide in Fig. 1(b), the main incident angles are still close to the normal directions of the cavity boundary. Then it is easy to know that most light will refract into free space and thus exclude the possibility of efficient coupling. Below we will show this intuitive picture is actually not accurate due to the presence of optical resonances.

We calculated the transmission spectra in the waveguide-disk system with Finite Element Method (FEM, Comsol Multiphsyics 3.5a). We set the numerical model as transverse magnetic (TM, E is perpendicular to the plane) polarization and the width of waveguide $w = 0.11R$. Here, the material dispersion is ignored. Figure 2(a) shows the normalized transmission spectrum taken at the plane marked as dashed line in Fig. 2(c). The transmittance at most frequencies is around 0.5, consistent with the intuitive picture and the calculation from Fresnel law. However, the transmittance at the normalized wavelength $kR \sim 5.187$ ($k = 2\pi/\lambda$, $\lambda$ is the wavelength) quickly drops to 0, clearly showing a counter-intuitive response. The corresponding field pattern in Fig. 2(b) reveals that the light is well coupled to a resonant mode with Azimuthal number $m = 10$ and radial number $l = 2$. Figure 2(c) shows the field pattern of mode at $kR \sim 5.233$ for a direct comparison. Associated with the high transmission, only a little field is confined within the cavity. Thus we can conclude that the resonance plays an important role in light trapping. What's more, it is worth pointing out that the higher order mode in Fig. 2 is not an exceptional case. Other resonance modes including fundamental mode can also be excited by end-fire injection.

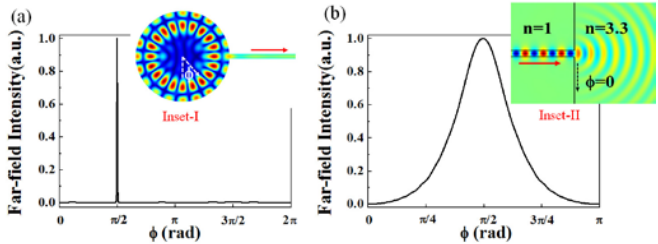

Fig. 1. (a) The far field pattern of the emissions from waveguide connected microdisk. Inset-I is the schematic picture of waveguide connected microdisk. (b) The angular distribution of the output beam from the waveguide as illustrated in the Inset-II.

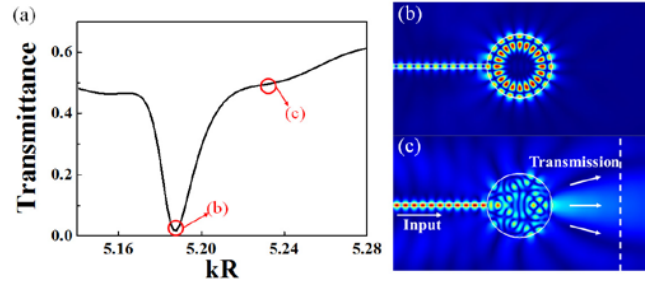

Fig. 2. (a) The transmission spectrum from $kR \sim 5.138$ to $kR \sim 5.28$. (b) and (c) are the corresponding far field patterns of modes at $kR \sim 5.185$ and $kR \sim 5.233$ marked in (a). The width of input waveguide is $0.11R$.

To estimate the coupling efficiency, we put a bus waveguide nearby the microdisk. Fig. 3(c) is the schematic picture, where four ports are clearly defined. We set the separation distance between bus waveguide and microdisk as $d$ and the waveguide width as $w$. The refractive index of waveguide is the same as microdisk with $n = 3.3$. The coupling efficiency is calculated by the ratio of $I_{WGM}/I_{input}$, where $I_{WGM}$ is the total light transmitting in WGMs of the microdisk and $I_{input}$ is the sum of incident light in the waveguide. As $I_{WGM}$ is hard to be determined, we placed a bus waveguide nearby the microdisk to couple out the light inside microdisk via evanescent waves. Due to the small coupling distance between microdisk and bus waveguide, only the circulating light along WGMs can be coupled into bus waveguide. By adjusting the separation distance, the coupling to bus waveguide can dominate the leakage. At the critical coupling point where the reflected and transmitted waves are both reduced to around 0 (see Fig. 3(b)), $I_{WGM}$ can thus be replaced by the sum of energy at port-2 and port-3 of the bus waveguide. Then the coupling efficiency can be estimated by $(I_{port-2} + I_{port-3})/I_{input}$.

Figure 3(a) shows the obtained spectra at port-2 and port-3 in a wide range of $kR$. Three peaks marked as 1-3 can be clearly found. Peak-1 is the same resonance as the one in Fig. 2(b). The total transmission (outputs in both port-2 and port-3) at peak-1 is as high as 75%. The detail information of peak-1 at four ports is shown in Fig. 3(b). Associated with the increasing of transmissions along ports 2 and 3, the intensities along port-4 (transmission to free space) and port-1 (reflection to input waveguide) reduce quickly. This means that most of the energy are directly coupled from waveguide to WGMs and transferred to the bus waveguide via the evanescent waves. Figure 3(c) shows the corresponding field pattern. Similar to the observation in Fig. 2, the high coupling efficiency is also associated with a resonant mode, indicating the importance of resonance again.

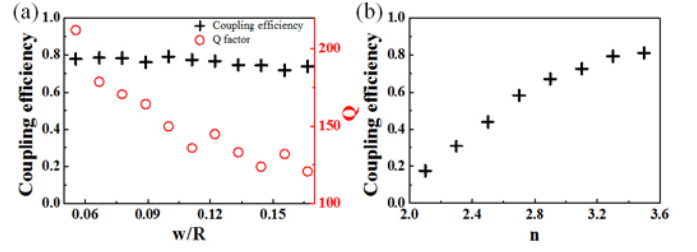

Fig. 4. (a) The coupling efficiency and Q factor as a function of waveguide width. (b) The coupling efficiency with different refractive index $n$.

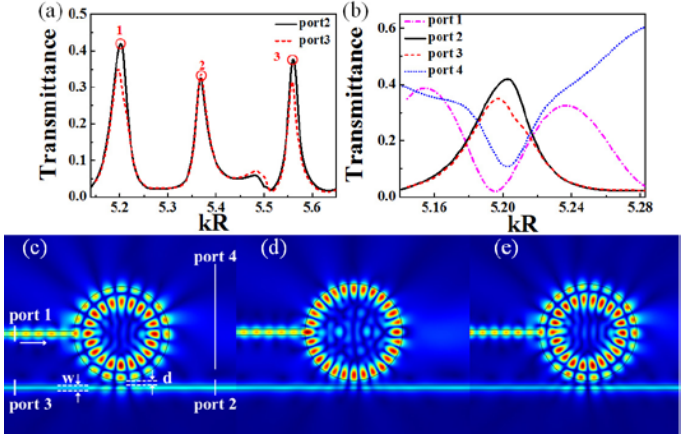

Fig. 3. (a) The transmission spectra at port-2 and port-3 of the waveguide connected microdisk. (b) The detailed spectra at all four ports, which are defined in (c). (c) - (e) are the corresponding electric field patterns of modes 1-3 in (a). The parameters $R$ and $w$ are the same as Fig. 2 and here $d = 0.089R$.

Besides the mode at $kR \sim 5.2$, we have also studied the direct coupling at different frequencies. The results are shown as peak-2 and peak-3 in Fig. 3(a). We can see that the transmittances of peak-2 and peak-3 along the bus waveguide are also as high as 62% and 70%, respectively. Figure 3(d) and (e) show their corresponding field patterns. More than the formation of WGMs at the peaks, we also observe that either the Azimuthal number or the radial number of two modes is different from peak-1. The peak in the middle of peak-2 and peak-3 is another resonance with Azimuthal number $m = 8$ and radial number $l = 3$. This mode is unconspicuous in the spectra for its relative lower transmittance. Thus we know that the end-fire injection is not a special phenomenon at one particular wavelength or particular set of resonances. Actually, it is quite generic for a number of resonances.

In additional to the insensitivity to the resonant frequencies, we also find that the end-fire injection is robust to the width of the input waveguide ($w$) and the refractive index of the microdisk ($n$). We take the coupling efficiency of mode-1 in Fig. 3 for an example. The corresponding results are summarized in Fig. 4. We can see that the coupling efficiency is kept at a high value around 70% - 80% when the width of waveguide is changed from $0.0556R$ to $0.1667R$ (see Fig. 4(a)). As such changes in $w$ are much larger than the fabrication tolerance in photolithography, this scheme is thus easily to be fulfilled in real experiment. The Q factors decrease as we enlarge $w$ for the increasing leakage from the waveguide. Meanwhile, we can see that the high coupling efficiency ($> 30\%$) can also be obtained from a wide range of refractive index from $n = 2.3$ to $n = 3.5$ (Fig. 4(b)). This shows that the end-fire injection is suitable for most semiconductors ranging from GaN for ultraviolet range to GaAs or Si for near infrared range where the tiny material dispersion is negligible in practical applications. Therefore, we can conclude that the high coupling efficiency can be easily realized in real experiment even though the cavity size is close to the resonant wavelengths [20].

Then the intriguing question is how the high coupling efficiency is formed. Figure 3(b) shows that the input waves are mainly separated into three parts, which are the transmission $E_T$ along port-4, the reflection $E_R$ along port-1, and the coupling to cavity $E_c$ recorded at port-2 and port-3. As $E_c$ will be affected by the microdisk and survive for some time, due to the wave effect, the input signals at later time can interfere with it. Once the constructive interference happens, $E_c$ will be increased. As the total intensity $I_{total} = I_T + I_R + I_c + I_S$ is kept as constant and $I = E^2$, the increase of field amplitude within microdisk will generate the enhancement in $I_c$ (also the coupling efficiency to microdisk) and the decreasing in the other parts. $I_S$ is the energy of surface scattering loss. We assume $I_S = 0$ in our model. Considering the field after interference is $E^2 = (E_c + E_{input})^2 = E_c^2 + E_{input}^2 + 2 \times E_c E_{input}$, the lifetime of the waves coupled to cavity turns to be essential. Once the light is coupled to the WGMs, the field amplitude $E_c$ can be accumulated to an extremely high value due to the relative long lifetime of the resonances. Then the constructive interference between the input waves and the WGMs can significantly improve the coupling from waveguide to disk (increase in $2 \times E_c E_{input}$) and suppress $I_{R,T}$ dramatically (see an example in Fig. 3(b)). Otherwise, the lifetime within the cavity is negligible and $E_c$ is also very tiny. Thus the transmission to port-4 (free space) is as high as expectation.

To verify our analysis, we have also calculated the same structure with finite difference time domain (FDTD) method, which gives consistent results with Figs. 2-4.

Moreover,

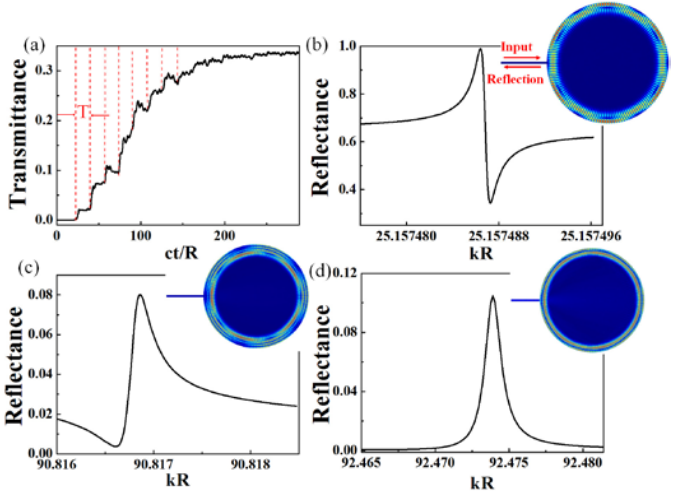

Fig. 5. (a) The time-dependent output of the mode of peak-1 in Fig. 3(a) at port-3. The results are calculated with FDTD method. c is the velocity of light in the microdisk. (b) The reflection spectra along the input waveguide around $kR \sim 25.1574865$. Here we set $w = 0.0375R$ and $n = 3.3$. (c) and (d) are the reflected spectra around $kR \sim 90.8168542$ and $92.473899$, respectively. Here, we set $w = 0.05R$ and $n = 1.45$. The insets in (b) - (d) are their corresponding field patterns at the reflection peaks.

FDTD can provide further information in time domain which is essential for understanding the counter-intuitive coupling. Figure 5(a) shows the time dependent output of mode-1 in Fig. 3(a) at port-3 as an example. We can see that the high transmission in port-3 requires time to build up, indicating the importance of the lifetimes. Most importantly, the build-up process increases step by step and the length of each step ($T$ in Fig. 5) is the same as the orbit length of the resonant mode. This clearly shows the accumulation of the energy in each resonant round and confirms that the constructive interference of the cavity mode and the end-fired injection. With the further increasing of time, the output from port-3 approaches to a saturation, where a balance between the input and the leakage happens.

The Q factors are relative low in Figs. 2, 3. This is mainly caused by the small cavity size [24]. It is worth to note that this coupling mechanism is not limited in the low Q resonances. This problem can be easily compensated by increasing the cavity size. Another approach is to introduce some special coupled modes with weak field distribution near the connected waveguide. The bus waveguide nearby the microdisk is nothing but a tool, which is used to calculate the coupling efficiency from connected waveguide to microcavity. In the following, we calculate the reflection spectra along the connected waveguide without the extra bus waveguide, as the end-fire injection mechanism is independent of it. As we have enlarged the microdisk size and excited a long-lived mode at $kR \sim 25.1574865$. Here we set $w = 0.0375R$, $n = 3.3$. The obtained high-Q mode is shown in Fig. 5(b) with $Q \sim 1.5 \times 10^7$, which is formed by mode coupling between $TM_{70,2}$ and $TM_{76,1}$ [18]. It is easy to see that the resonant mode has a hexagonal-shape field pattern and is confined along a period-6 orbit. As the hexagonal-shape resonance has weak field distribution near the connecting waveguide, the waveguide thus will not affect the total internal reflection within the coupled mode directly, making the high Q factor possible. More than the high-refractive-index microdisk, the high-Q modes can be excited in a lower-refractive index microcavity. Fig. 5(c) and (d) are the obtained reflection spectra along the input waveguide. The refractive index of microcavity and waveguide is $n = 1.45$ (for most of the microspheres and microtoroids) and the waveguide width is $w = 0.05R$. Similar with Fig. 5(b), a pentagonal-shape (Fig. 5(c)) and a heptagon-shape (Fig. 5(d)) mode can be excited with $Q \sim 4.9 \times 10^4$ and $\sim 6.9 \times 10^4$, respectively. Although the coupling efficiency decreases, the change in reflection intensity ($\sim 8\%$ and $\sim 10\%$) makes the sensors applicable to real experiments. Interestingly, as most of the microspheres are fabricated by heating the end tips of tapered fibers, the end-fire injection mechanism can thus efficient coupling light into them. Then the relative long-lived resonances along the "longitude" direction can be utilized for optical sensing. Most importantly, this kind of sensors can be naturally compatible with the fiber communication system through the optical fiber and be used in long distance sensing.

## 3. Conclusions

In summary, we numerically studied the transmission characteristics of the structure microdisk connected with an input waveguide. Counter to our intuitive thought, we show that more than 75% light can be coupled into the microdisk due to the interference between the input light and resonant modes. This coupling method is found to be robust to the width of input waveguide and the refractive index, making it suitable for practical applications.

**Acknowledgments** This work is supported by NSFC11204055, NSFC61222507, NSFC11374078, NCET-11-0809, Shenzhen Peacock plan under the Nos. KQCX2012080709143322 and KQCX20130627094615410, and Shenzhen Fundamental research projects under the Nos. JCYJ20130329155148184, JCYJ20140417172417110, JCYJ20140417172417096.